\documentstyle[12pt]{article}
\newcommand{\bkg}{\mbox{$B\to K^\ast \gamma$}}
\newcommand{\bkpg}{\mbox{$B\to K^\ast + \gamma$}}
\newcommand{\bskpg}{\mbox{$B_s\to \phi + \gamma$}}

\newcommand{\lsim}{\mathrel{\lower4pt\hbox{$=$}}
\hskip-10pt\raise1.6pt\hbox{$<$}\;}

\newcommand{\ltsim}{\mathrel{\lower4pt\hbox{$\sim$}}
\hskip-12pt\raise1.6pt\hbox{$<$}\;}

\newcommand{\gtsim}{\mathrel{\lower4pt\hbox{$\sim$}}
\hskip-12pt\raise1.6pt\hbox{$>$}\;}

\begin{document}

\begin{titlepage}
\noindent Wash.\ U.\ HEP/93-35 \hfill November, 1993\\
\begin{center}

{\Large\bf \bkpg\ and \bskpg\  on the Lattice}

\vspace{1.5cm}
{\bf C. Bernard$^{(1)}$, P. Hsieh$^{(2)}$ and A. Soni$^{(3)}$}\\
\end{center}
\vspace{1.0cm}
\begin{flushleft}
{}~~$^{(1)}$Physics Department, Washington University, St. Louis, MO\ \
63130, USA \\

{}~~$^{(2)}$Smithsonian Astrophysical Observatory, Cambridge, MA\ \ 02138 \\

{}~~$^{(3)}$Physics Department, Brookhaven National Laboratory, Upton,
NY\ \ 11973, USA \\
\end{flushleft}
\vspace{1.5cm}

\abstract{
A lattice calculation of the form factors that determine the
``hadron\-i\-zation ratios'', such as $R_{K^\ast}$ and
$R_{\phi}$, where $R_{K^\ast} \equiv [\Gamma (\bkg)/\Gamma(b
\to s \gamma)]$, is presented in the quenched approximation.
Lattice data shows strong evidence for the scaling law
suggested by heavy quark symmetry for
one of the form factors ({\it i.e.}\ $T_2$).
The data also gives strong support for the simple pole ansatz
for the $q^2$ dependence of $T_2$ in the range of available $q^2$.
We thus find $T_2(0) = .10 \pm.01\pm.03$ yielding $R_{K^\ast}
=(6.0\pm1.2\pm3.4)\%$; we also find $R_{\phi}=(6.6\pm1.3\pm3.7)\%$.
}
\vfill
\end{titlepage}

\baselineskip 30truept
The loop decays of the $b$-quark have long been noted for their capacity
to provide important tests of the Standard Model (SM) \cite{1}. Since
many of  these
decays are short distance dominated at the quark level, their inclusive
rates are amenable to perturbation theory. Thus for inclusive processes
reliable predictions can be made. The fact that the $b$-quark has a
relatively long lifetime facilitates experimental tests of the
theoretical predictions. In particular, in the SM, the simple decay $b
\to s + \gamma$ is predicted to have a branching ratio which varies
from $\simeq 2\times 10^{-4}$ to $\simeq 4\times 10^{-4}$ as the top quark mass
varies from
100 to 200 GeV \cite{2}.
Assuming three generation unitarity,
along with
$V_{cs} \simeq V_{tb} \simeq 1$, it is easy to see that $b\to s+
\gamma$ is  independent of CKM angles \cite{3}
to a very good approximation.
Furthermore,
$b \to s + \gamma$ is also noted for its sensitivity to extensions
of the SM \cite{4}. Finally, we note that $b \to s(d) + \gamma$ can also
lead to a determination of $V_{ts}$ and $V_{td}$. However, the full
potential of the loop decays of the $b$-quark is very difficult to
capitalize upon unless reliable theoretical predictions can also
be made at the exclusive level.

 Indeed, the inclusive process $b \to s + \gamma$ is
challenging to measure experimentally; whereas a corresponding exclusive
mode ({\it i.e.}, \bkpg) has a distinctive signature and is  much more
accessible to experiment.  Thus a meaningful confrontation between
experiment and the underlying  electroweak theory can be facilitated
through  a knowledge of the ``hadronization ratio'', $R_{K^\ast}$:

\begin{equation}
R_{K^\ast} \equiv \frac{\Gamma(\bkg)}{\Gamma
(b\to s \gamma)}
\end{equation}

\noindent which is the probability for the formation of the $K^\ast$.
The evaluation of this ratio by continuum methods has proven to be
extremely difficult. This is reflected in the wide range $\sim 1$
to $\sim97\%$ in the value of $R_{K^\ast}$, as calculated by quark
models, QCD sum rules, heavy quark symmetry (HQS) extended to include the
$s$-quark, {\it etc}.\ (see Table~1) \cite{5}. Under the circumstances, the
hard earned experimental determination of the branching ratio for
\bkpg\ may  only be used
to select amongst various models of hadronization rather than to test
the underlying theory. It is thus clearly important to explore
the use of lattice methods for treating such exclusive decays of $B$ mesons.

At the quark level
the decay is described by an effective  Hamiltonian \cite{2,5}:

\begin{equation}
H_{eff} = G_{g_1g_2g_3}(m_t,\mu)\  V_{ts}\  \bar s(x) \sigma_{\mu\nu} b_R(x)\
F^{\mu\nu}(x)
\end{equation}

\noindent where $F^{\mu\nu}$ is the photon field strength tensor,
$b_R \equiv \frac{1+\gamma_5}{2} b$, and
the $c$ number coefficient $G_{g_1g_2g_3}(m_t,\mu)$ depends on
all the three gauge couplings of the Standard Model (SM),
the mass ($m_t$) of the top quark and a renormalization point $\mu$. The
dependence on the CKM angle $V_{ts}$ has also been factored out.
$V_{tb}$ is assumed
to be
1, and the negligibly small $u$ quark contribution
is ignored.

As usual \cite{6} the lattice is used for a non-perturbative
evaluation of the matrix element $M_\mu \equiv \langle V(k)|J_\mu|P(p)
\rangle$, where $P$ is the
initial pseudoscalar heavy-light meson, $V$ is the final vector
meson,
$J^\mu\equiv \bar s \sigma^{\mu\nu} q_\nu
b_R =
(v + a)_{\mu}$ is the current,  with
${\it v_{\mu}}$ and ${\it a_{\mu}}$ the vector and axial parts,
and $q\equiv p-k$ is the
4-momentum of the photon.
In general, the Euclidean matrix element can be
parameterized in terms of three form factors \cite{5,7}:

\begin{eqnarray}
M_\mu & = & 2\epsilon_{\mu\nu\lambda\sigma}\eta^\nu(k) p^\lambda k^\sigma
T_1(q^2) + [ \eta_\mu(k) (m^2_H-m^2_V) \nonumber \\
&& -\eta\cdot q(p+k)_\mu]T_2(q^2) +\eta\cdot q\left[q_\mu-
\frac{q^2}{m^2_H-m^2_V} (p+k)_\mu \right] T_3(q^2)
\end{eqnarray}

\noindent (Our $\gamma$ matrices obey
$\{\gamma_\mu, \gamma_\nu\} = 2\delta_{\mu\nu}$, and momenta
are defined by $p_\mu = (E, i\vec p)$, so that $p^2=m^2$ on shell.)
For a lattice calculation  it is
simpler to note that the $T_1(q^2)$ term arises purely from the vector
piece of $J_\mu$ and the $T_2$ and $T_3$ terms given above arise
from the axial piece. The third term does not contribute when the
photon is on shell. Furthermore, at the end-point, ($q^2=q^2_{max}
\equiv(m_H-m_V)^2$), where the final and initial mesons are both at
rest, $T_3$ term does not contribute to the axial matrix element.
Since also at that kinematic point no momentum injection
is required,
$T_2(q^2_{max})$ can be readily, and rather cleanly,
evaluated on the lattice. Although $q^2=0$ (the point of direct
physical interest) is not exactly accessible to the
lattice, in many instances the parameters used in the current simulation
do allow $q^2$ to be extremely
small, {\it i.e.}\ $q^2/m^2_H \le.1$\null. Finally
we note that using the identity $\sigma_{\mu\nu} \gamma_5 \equiv
-\frac{1}{2} \epsilon_{\mu\nu\lambda\rho} \sigma^{\lambda\rho}$ one
can show that

\begin{equation}
T_2(0) = T_1(0)\ .
\end{equation}

Now the hadronization ratio of interest takes the simple form
(for $m_s \ll m_b$):

\begin{equation}
 R_{K^\ast} = 4\left(\frac{m_B}{m_b}\right)^3 \left[
1-\frac{m^{\ast2}_K}{m^2_B} \right]^3 |T_1(0)|^2 \ .
\end{equation}

\noindent On current lattices $q^2=0$ (or near that point) is
inaccessible for very heavy meson masses, say $m_H\ge3.5$
GeV\null. So at $m_H\sim m_B$, $T_1(0)$ is not directly calculable.
However
HQS \cite{8} allows one to predict the behavior of
$T_2(q^2_{max})$ at large $m_H$. Indeed, when $q^2=q^2_{max}$ no large
momenta is transferred to the recoiling light hadron, so
a straightforward argument shows
that $\sqrt{m_H} T_2(q^2_{max}) \to {\it const.} $ (up to logarithms)
as $m_H \to \infty$.
This makes possible a controlled extrapolation
of $T_2(q^2_{max})$.
Our strategy on the lattice will thus be:

1) test pole dominance of $T_2$ at fixed $m_H$, to the extent that the
data allow, by deducing $T_2(0)$ from $T_2(q^2_{max})$ using the equation:
\begin{equation}
T_2(0) = T_2(q^2_{max}) \left[ 1- \frac{q^2_{max}}{m^2_H} \right]
\end{equation}
\noindent and comparing to $T_1(0)$ using eq.\ (4).  $T_1(0)$ is also
obtained using pole dominance, but only from $T_1$ at small values of $q^2$
($q^2/m_H^2 < 0.1$).
Pole dominance does not appear to work well for $T_1(q^2)$ with large
$q^2$.

2) extract
$T_2(q^2_{max})$ at $m_H=m_B$ by fitting the data to the form suggested by
HQS, namely:

\begin{equation}
\sqrt{m_H}\; T_2(q^2_{max}) = A_1+A_2\left(\frac{1}{m_H}\right) \ .
\end{equation}

3) use pole dominance for $T_2$ at $m_H=m_B$ to deduce $T_1(0)=T_2(0)$
from $T_2(q^2_{max})$.

We remark that in testing pole dominance, we have simply used the
pseudoscalar mass in eq.(6) because in the limit of large $m_H$,
HQS implies that resonances of different spin parities
become degenerate \cite{8}.
Note also that step 3) uses pole dominance over a wider range of $q^2$
($q^2_{max}/m_B^2 \approx 0.65$) than can be explicitly checked
in step 1)  ($q^2_{max}/m_H^2 \ltsim 0.3$).  We attempt to estimate below
the systematic error associated with this step.

We mention the following technical points, regarding the lattice
calculations, in brief \cite{9}.
\medskip

\noindent 1) The recently proposed normalization of the Wilson quarks
on the lattice \cite{10,11,12}:

\begin{equation}
\psi^{\rm continuum} = \sqrt{2 \tilde\kappa\exp(a\tilde m)} \psi^{\rm lattice}
\end{equation}

\noindent where

\[ a\tilde m= ln \left[ \frac{1}{2\tilde\kappa} - 3 \right] \]

\noindent and $\tilde\kappa = \kappa/8\kappa_c$ ($\kappa_c$ is the
critical hopping parameter) is used. Thus
the leading corrections that become important as $am$ gets large are
automatically included.
\medskip

\noindent 2) For the renormalization of the tensor current we
incorporate the correction calculated in lattice weak coupling
perturbation theory to one loop order \cite{13}. However, following
Lepage and MacKenzie \cite{11}, the tadpole contribution is removed
from the
correction (it is already included
in eq.\ (8)), and a ``boosted'' value of
$g_3 = g_V(1/a)$ is used.

We have done the calculation of $T_1$ and $T_2$ on four different sets
of lattices: A) $\beta=6.3$, $24^3\times61$ (20 configurations,
$a^{-1}=3.01$ GeV), B1) $\beta=6.0$, $24^3\times39$ (8 configurations,
$a^{-1}=2.29$ GeV), B2) $\beta=6.0$, $24^3\times39$ (a second set of 8
configurations, $a^{-1}=2.29$ GeV),  and C) $\beta=6.0$, $16^3\times39$
(19 configurations, $a^{-1}=2.10$ GeV).
The lattice spacings given above are determined
through a calculation of $f_\pi$ with the same point sources
that are used here \cite{12}.
The ``$B$'' is always taken at rest; the ``$K^*$''
is given lattice momentum $(0,0,0)$, $(1,0,0)$, $(1,1,0)$ or
$(2,0,0)$, with $(2,0,0)$ used only on B1 and B2.
Preliminary results of this computation have been presented
previously. \cite{14}

We first work in the case when the masses of the two light quarks are
held equal. Experimentally this situation corresponds to the decay,
for example, \bskpg\ \null.
For the light quark  we use $\kappa=.148$  at $\beta=6.3$
and
$\kappa=.152$ at $\beta=6.0$,
yielding a vector
meson in the final state with mass $\approx 1.3$ GeV\null. The
dependence of the amplitude on the heavy quark mass is then
studied. Specifically, for $\beta=6.3$, we use $\kappa=0.140$, 0.125, 0.110
and 0.100 for the heavy quark.
Results are given in Table 2; the last column shows that
$\sqrt{m_H}\;T_2(q^2_{max})$ is approximately constant.
We then fit the data to the two parameter form (eq.\ (6)) suggested by HQS
taking the correlations in the data into account through covariant fits.
For the $\beta=6.3$ data we find $A_1=.806\pm.069\
({\rm GeV})^{1\over2}$, $A_2=-.545\pm.082\ ({\rm GeV})^{3\over2}$,
$(\chi^2/dof \approx 2.3/2)$. Thus

\[T_2(m_H=m_{B_s},q^2_{max}) = .304\pm.030 \ . \]

We now discuss the systematic errors on $T_2$, first
considering those relevant to \bskpg.
To correct for the physical s-quark
we also study the matrix elements with $\kappa =0.149$
(corresponding to vector meson of mass about 1.1 GeV), at $\beta = 6.3$.
We find a shift in $T_2$, from its value at $\kappa = 0.148$ of
about $-7\%$.
Extrapolating to the physical s-quark would give a shift
of $-10\%$.
In passing we mention that a similar study of our lattices
at $\beta = 6.0$ indicates a smaller error than the 10\% seen at
$\beta=6.3$.

We now assess the  systematic error due to the use of
heavy quarks with $am \gtsim 1$.
For that purpose, we fit to the two parameter form
using the two lightest quarks from our heavy set (of four)
at $\beta = 6.3$; {\it i.e.}\ we retain only $\kappa = 0.140$ and $\kappa
= 0.125$. We find a shift in the value of $T_2$ of 3.1\%.

To estimate scale breaking errors we compare the fit for the $\beta=6.3$
data with the heavy quarks at $\kappa=.140$ and .125 to the fit for the
$\beta=6.0$ data with the corresponding heavy quarks at $\kappa=.135$
and .118. We attribute the difference of 12.2\% to scaling
violations.

The systematic errors due to finite size effects are deduced by comparing
the value of $T_2$ on our $16^3$ lattice with the one on the
$24^3$ lattice, both at $\beta = 6.0$. We find
a difference of 9.4\%.

Adding in quadrature the errors due to the four sources mentioned above we find
a
total systematic error of 19\%.
In passing we note, however, that
the systematic error due to each of these four sources is actually
smaller than the statistical errors in the appropriate subset of data.
It is, therefore, quite likely that the estimate of 19\% is a
conservative one. Thus, we arrive at:

\begin{equation}
T_2(m_H=m_{B_s},q^2_{max}) = .304\pm.030\pm.057 \ . \end{equation}

Table~3 summarizes our test of the pole dominance for $T_2$.
By examining the agreement between $T_2(0)$ and $T_1(0)$ we see that,
within the available range of $q^2_{max}/m^2_H\le0.3$, the pole-model seems
to work very well. We must note, however, that in the actual physical
reactions of interest $q^2_{max}/m^2_B$ approaches about
0.65. To estimate the error involved, we note that the biggest
difference between $T_1(0)$ and $T_2(0)$ is $\sim9\%$ (for lattice C).
Scaling by the increased range in $q^2$ for the physical reaction
($0.3\to 0.65 $), we arrive at an error of $20\%$.
Since the
data points with higher $q^2_{max}/m^2_H$ in Table~3 seem to
support pole ansatz just as well as those with lower values of
$q^2_{max}/m^2_H$,
this is likely to be an overestimate, but we wish to be
conservative.
Using eqs.(4), (5),
(6) and (9) we thus find:

\begin{equation}
T^{B_s \to \phi}_1(0) = T^{B_s \to \phi}_2(0) = 0.104\pm0.010\pm0.028 \ .
\end{equation}

\begin{equation} R_{\phi} = (6.6\pm1.3\pm3.7)\% \ ,  \end{equation}

\noindent which is the hadronization ratio for \bskpg. Note that in this
calculation we have taken $m_b = 4.5$ GeV, so that we
may use the result for BR($b\to s\gamma$) given by
Misiak \cite{2}. A 13\% uncertainty is added
in quadrature to the systematic errors on $R$ corresponding to an
assumed 200 MeV uncertainty in $m_b$.

Next we turn our attention to \bkpg.
For this purpose we study matrix elements with unequal masses for the
light quarks. For example, at $\beta = 6.0$ we take the spectator quark
with $\kappa =0.154$ and the ``s'' quark (corresponding
to the light quark that results from the weak decay of the $b$-quark)
with $\kappa = 0.152$. Furthermore, we have to extrapolate in the
masses of these two quarks. In particular, the spectator quark
requires extrapolation to the chiral limit ({\it i.e.} $\kappa_{c} = 0.157$
at $\beta = 6.0 $). For this study we use the $\beta = 6.0$,
$24^3$ lattice as it has the largest physical volume. This lattice
has two independent sets of configurations with eight configurations
in each sample. As a result we find that extrapolation to the physical
limit causes a shift in $T_2$ from its value calculated with
degenerate light quarks ( $\kappa = 0.152$ at $\beta = 6.0$) of
about $+7\%$. We shift the central value of $T_2$ accordingly and also
include this additional 7\% in the systematic errors.
Consequently we arrive at:

\begin{equation}
T_2(m_H=m_B,q^2_{max}) = .325\pm.033\pm.065 \ . \ \end{equation}

Once again we use pole dominance to get:
\begin{equation}
T^{B \to K^*}_1(0) = T^{B \to K^*}_2(0) = 0.101\pm0.010\pm 0.028 \  ,
\end{equation}

\begin{equation}
R_{K^*} = ( 6.0\pm1.2\pm3.4 )\% \ .
\end{equation}

Now, as mentioned earlier, the inclusive branching ratio for
$b \to s \gamma$ is predicted to lie in the range of
about $(2- 4)\times10^{-4}$ depending on $m_t$. Thus, in the SM, there is
a bound, $BR(b \to s + \gamma) \leq 4 \times 10^{-4}$,
corresponding to $m_t \approx 200$ GeV\null. Combining this upper
bound with the above lattice result one gets:

\begin{equation}
BR(B \to K^* \gamma) \ltsim (2.4 \pm0.5\pm1.4) \times10^{-5}  \ .
\end{equation}
We recall now the recent CLEO result \cite{15}:
\begin{equation}
BR(B \to K^* \gamma) = (4.5 \pm1.5 \pm0.9) \times10^{-5}\ .
\end{equation}

Given the size of the errors in the lattice calculation,
as well as in the experiment, the CLEO result is
certainly {\it not inconsistent} with the expectations based on the lattice.
However, we note that the numbers seem to mildly favor a rather heavy
top quark.

In an attempt
to quantify the statement about $m_t$ we note that the experimental
result along with the lattice implies:

\[BR(b \to s \gamma) = (4.5 \pm1.5 \pm0.9)\times 10^{-5}/
(6.0 \pm1.2
\pm3.4) \times10^{-2} \null \approx (7.5 \pm 5.4) \times 10^{-4} \]

\noindent where we have assigned a $\sim70\%$ combined error to the lattice
plus
the experimental result. At the 1-$\sigma$ level
one then finds
$m_t \gtsim 100 GeV$. However  modest improvements in the the lattice
and/or experimental results could produce a rather stringent bound.

To summarize, we have used lattice methods to evaluate the form factors
for the radiative $B$ transitions. We emphasize that the heavy quark limit
of QCD \cite{8} enters in an important way in making this calculation
feasible on current lattices. We also want to highlight two drawbacks of the
present effort. First, numerical limitations did not allow us
to check pole dominance for the specific value of the momentum transfer
relevant to the experiment. However, lattice parameters did allow us to
check the pole model for $T_2(0)$
for an appreciable range of momenta, giving strong
support to its validity. We have included what we believe
is a conservative estimate of $20\%$ systematic error
due to the use of pole dominance.
The second limitation is, of course, the quenched
approximation. It is generally believed that with the use of a
physical
quantity ({\it e.g.} $f_\pi$ in our work \cite{12}) to set the scale for the
lattice calculations, errors due to quenching are likely to be quite
small, perhaps $\leq 10\%$, in the form factors of interest
here.
It is, therefore, unlikely that the present limitations would
seriously affect our results, given the $\sim28\%$ error in amplitude.
Quenched simulations are now in progress
that should allow us to improve the calculations to the 10--15\%
level. At that stage errors due to quenching may also start to
become relevant.
\bigskip

\noindent{\it Acknowledgements}

We thank D. Atwood, N. Deshpande, M.\ Danilov,
J. Hewett, M. Misiak,
D.\ Richards, Y. Rozen, H.\ Shanahan
and E. Thorndike
for discussions.
C.B.\ was partially supported by the DOE under grant number
DE2FG02-91ER40628; P.H.\ and A.S.\ by the
DOE grant number DE-AC0276CH00016. The computing for this project was
done at the National Energy Research Supercomputer Center and at the
San Diego Supercomputer Center.

\pagebreak

\begin{quotation}
Table 1: A sample compilation of the predictions for $R_{K^\ast} \equiv
[ \Gamma (B-K^\ast\gamma)/\Gamma (b\to s\gamma)]$. See Ref.~5.
\end{quotation}
\begin{center}
\begin{tabular}{lc}
Author(s) & $R_{K^\ast}$ \\ \\
O'Donnell (1986) & 97\% \\
Deshpande {\it et al}. (1988) & $6\%$ \\
Domingues {\it et al}. (1988) & $28\pm11\%$ \\
Altomari (1988) & 4.5\% \\
Deshpande {\it et al}. (1989) & 6--14\% \\
Aliev {\it et al}. (1990) & 39\% \\
Ali {\it et al}. (1991) & 28--40\% \\
Du {\it et al}. (1992) & 69\% \\
Faustov {\it et al}. (1992) & 6.5\% \\
El-Hassan {\it et al}. (1992) & $\sim0.7\%-12\%$ \\
O'Donnell {\it et al}. (1993) & $\sim10\%$ \\
Ali {\it et al}. (1993) & $13\pm3\%$ \\
Ball (1994) & $20\pm6\%$ \\ \\

This work & $(6.0\pm1.2\pm3.4)\%$ \\
\end{tabular}
\end{center}

\pagebreak

\begin{quotation}
Table 2: Lattice data on four sets of lattices. $\kappa_1$
represents the intial heavy quark undergoing weak decay,
$\kappa_2$ the light quark emerging from the weak decay.
The spectator quark is taken to have $\kappa_2$ as well.
$m_H$ and $m_V$ are the masses of the initial and the final
$0^-$ and $1^-$ mesons respectively and
$r_{max} \equiv [q^2_{max}/m_H^2]$.

\end{quotation}
\begin{center}
\begin{tabular}{l|c|c|r|c|r|c|c}
\hline
$\beta(a^{-1}/\mbox{GeV})$ & Lattice & $\kappa_1, \kappa_2$  & $am_H$ &
$am_V$  & $r_{max}$ & $T_2(q^2_{max})$ & $\sqrt{m_H}T_2(q^2_{max})$ \\
$\{\kappa_c\}$ & Set &&&&&& $({\rm GeV})^{1\over2}$ \\
\hline
6.3(3.01) & A & 140,148 & .590 & .422 & .081 & $.406\pm.046$  &
$.54\pm.06$ \\
$\{0.151\}$ &&&&&&& \\
 & A & 125,148 & .934 & .422 & .301 & $.384\pm.044$ & $.64\pm.08$ \\
 & A & 110,148 & 1.248 & .422 & .443 & $.364\pm.048$ & $.71\pm.10$ \\
 & A & 100,148 & 1.465 & .422 & .508 & $.346\pm.055$ & $.73\pm.12$ \\
 &&&&&&& \\
6.0(2.29) & B1 & 135,152 & .894 & .561  & .139 & $.409\pm.090$  &
$.58\pm.13$ \\
$\{0.157\}$ &&&&&&& \\
 & B1 & 118,152 & 1.244 & .561 & .301 & $.371\pm.105$ &
$.63\pm.18$ \\
  &&&&&&& \\
6.0(2.29) & B2 & 135,152 & .891 & .566  & .139 & $.478\pm.090$  &
$.69\pm.13$ \\
 & B2 & 118,152 & 1.245 & .566 & .301 & $.415\pm.105$ &
$.71\pm.17$ \\
  &&&&&&& \\
6.0(2.10) & C & 142,152 & .734 & .564 &  .053 & $.470\pm.062$
    & $.58\pm.08$ \\
 & C & 135,152 & .888 & .564 &  .133 & $.459\pm.065$
    & $.63\pm.09$ \\
 & C & 118,152 & 1.241 & .564 & .298 & $.414\pm.089$
& $.67\pm.14$ \\
\end{tabular}
\end{center}

\bigskip\bigskip

\pagebreak

\begin{quotation}
Table 3: Test of the pole model for the $q^2$ dependence of the
form factors;
in particular, that of $T_2$. $T_2(0)$ and $T_1(0)$ are deduced,
from $T_1(q^2)$ and $T_2(q^2)$ seen on the lattice, by using pole
dominance {\it i.e.} eq.\ (6). Note $r\equiv q^2/m^2_H$.
\end{quotation}
\begin{center}
\begin{tabular}{l|c|c|c|c|c|c|c}
\hline
Lattice & $\kappa_1,\kappa_2$  & $r$ &
 $r_{max}$ & $T_1(q^2)$ & $T_2(q^2_{max})$ & $T_1(0)$ & $T_2(0)$ \\
Set &&&&&&& \\
\hline
A & 125,148 & $.002$ & .300 & $.259\pm.035$ & $.384\pm.044$ & $.259\pm.035$ &
$.269\pm.032$ \\
&&&&&&& \\
B1 & 135,152 & $.009$ & .139 & $.391\pm.069$ & $.409\pm.090$ & $.388\pm.068$ &
$.352\pm.077$ \\
&&&&&&& \\
B2 & 135,152 & $.009$ & .139 & $.436\pm.092$ & $.478\pm.090$ & $.432\pm.091$ &
$.411\pm.078$ \\
&&&&&&& \\
B1 & 118,152 & $-.034$ & .301 & $.264\pm.050$ & $.371\pm.105$ & $.272\pm.051$ &
$.260\pm.073$ \\
&&&&&&& \\
B2 & 118,152 & $-.034$ & .301 & $.316\pm.110$ & $.415\pm.100$ & $.327\pm.113$ &
$.290\pm.070$ \\
&&&&&&& \\
C & 118,152 & $-.069$ & .298 & $.300\pm.039$ & $.414\pm.089$ & $.321\pm.042$
&$.291\pm.062$ \\
&&&&&&& \\
\end{tabular}
\end{center}


\begin{thebibliography}{99}
\bibitem{9} More details of this work will be given in a
forthcoming article.
\end{thebibliography}
\end{document}